 \documentclass[11pt,twocolumn]{article}

 \pdfoutput=1

 \usepackage[english]{babel}
\usepackage{graphicx} 
\usepackage{amsmath}

\usepackage{enumitem}

\usepackage{float}
\usepackage{ifthen}
\usepackage{hyperref}

\usepackage{titlesec}

\titlespacing*{\section}{0pt}{0.2\baselineskip}{\baselineskip}

\begin{document}

\pagestyle{empty}


\twocolumn[
\begin{center}
{\bf \huge On consequences of measurements of turbulent}\\
\vspace*{1mm}
{\bf \huge Lewis number from observations.}\\
\vspace*{3mm}
{\Large \bf by Pascal Marquet${}^{\star}$, William Maurel${}^{\dag}$ and 
Rachel Honnert${}^{\star}$} \\
\vspace*{3mm}
{\Large (WGNE Blue-Book 2017)}. \\
\vspace*{2mm}
{M\'et\'eo-France. ${}^{\star}$CNRM/GMAP. ${}^{\dag}$CNRM/GMEI.
 Toulouse. France.}
{\it E-mail: pascal.marquet@meteo.fr} \\
\vspace*{1mm}
\end{center}
]





 \section{\underline{\Large Motivations}} 
\vspace{-4mm}

Almost all parameterizations of turbulence in NWP models and GCM make the assumption of equality of exchange coefficients $K_h$ for heat and $K_w$ water.
These two exchange coefficients are applied to the two moist-air Betts  (1973) ``conservative'' variables
\vspace{-3mm}
\begin{align}
 \theta_l & \;  = \; \; 
     \theta \: \exp \left( - \:
         \frac{L_v \: q_l \: + \: L_s \: q_i}{c_{pd} \: T} 
              \right)
\label{equ_thl} \: , \\
 q_t & \; = \; \;   q_v \: + \: q_l \: + \: q_i
\label{eq_qt} \:
\end{align}

\vspace{-4mm}
\noindent 
(where $\theta = T \: (p_0/p)^{R_d/c_{pd}}$ is the dry air potential temperature)
to compute the vertical turbulent fluxes written as:
$\overline{w'\, \theta'_l} = - \, K_h \: \partial \overline{\theta_l}/\partial z$ and
$\overline{w'\, q'_t} = - \, K_w \: \partial \overline{q_t}/\partial z$.

However, large uncertainties exists in old papers published in the 1950s, 1960s and 1970s, where the turbulent Lewis number $\mbox{Le}_{t} = K_h / K_w$ have been evaluated from observations.
Some papers are favourable to the hypothesis $K_h = K_w$ and $\mbox{Le}_{t}=1$, while others have observed higher values, up to $\mbox{Le}_{t} > 4$.

Moreover, the use of the Betts variable $\theta_l$ is based on an approximate moist-air entropy equation and this formulation has been improved in Hauf and H\"oller (1987) and  Marquet (2011, 2015), where the new potential temperature $\theta_s$ is defined as synonymous of the moist-air entropy.

The aim of this note is: 
1) to trust the recommendations of Richardson (1919), who suggested to use the moist-air entropy as a variable on which the turbulence is acting;
2) then to replace $\theta_l$ by the third-law entropy value $\theta_s$, which must correspond to a new exchange coefficients $K_s$;
3) compute $K_s$ and $\mbox{Le}_{ts} = K_s / K_w$ from observations (M\'et\'eopole-Flux and Cabauw masts) and from LES and SCM outputs for the IHOP case (Couvreux {\it et al.\/}, 2005).

\vspace{-0.5mm}
 \section{\underline{\Large The moist-air entropy flux}} 
\vspace{-4mm}

The specific (per unit mass of {\it moist-air\/}) entropy is defined in Marquet (2011, 2015) by 
$s  =  s_{ref}  + c_{pd} \: \ln(\theta_{s})$, 
where $s_{ref}$ and $c_{pd}$ are two constants.
If liquid water or ice do not exist, $\theta_l=\theta$ and the first-order approximation of the moist-air entropy potential temperature is 
${\theta}_{s} \approx \theta \: \exp(\Lambda \: q_v)$, 
where  $\Lambda \approx 5.87$ is a constant which depends on the third-law reference values of entropy of dry air and water vapour.
The second-order approximation derived in Marquet (2016)  writes
\vspace{-2.mm}
\begin{align}
  \theta_s  & \, \approx \; 
   \theta \; \exp(\Lambda \: q_v) \: \exp[ \: - \: \gamma \: \ln(r_v/r_{\ast}) \: q_v \: ]
\label{eq_ths2} \: ,
\end{align}
where $\gamma \approx 0.46$ and $r_{\ast} \approx 12.4$~g/kg are two constants.

With Reynolds hypotheses, the flux of moist-air entropy potential temperature can be written as
\vspace{-2mm}
\begin{align}
\! \!
  \overline{w' \theta'_s}  & \, = \; 
   - \: K_s \: \frac{\partial \overline{\theta_s}}{\partial z}
\label{equ_CS} \: , \\
\! \!\overline{w' \theta'_s} & \;  \approx \;  \;
\exp(\Lambda \: \overline{q_v}) \; \;  \overline{w' \theta'} 
 \nonumber \\
   &  \; \;  \: + \:
\overline{\theta_s} \;   \left[ \: \Lambda 
           - \gamma \: \ln\!\left(\frac{\overline{r_v}}{r_{\ast}}\right)
           - \frac{\gamma}{1-\overline{q_v}} \; \right]
 \;  \overline{w' q'_v} 
\label{eq_wths} \: .
\end{align}

\vspace{-3mm}
\noindent 
This flux is a weighted sum of the fluxes for $\theta$ and $q_v$.
And if the turbulence is to be represented by the flux of $\theta_s$ and $q_v$, the corresponding flux of $\theta$ is given by 
\vspace{-2mm}
\begin{align}
 \!
  & \overline{w' \theta'}
  \, \approx
   \; - \: K_w \: \;
   \mbox{Le}_{ts} \; \; \frac{\partial \overline{\theta}}{\partial z}
     \label{equ_flux_ths_Lets} \\
 \! \! \! \! \! \! \!
  &  - K_w \;
   \left( \mbox{Le}_{ts} - 1 \right) \:
   \left[ \: \Lambda - \gamma \: \ln\!\left(\frac{\overline{r_v}}{r_{\ast}}\right)
           - \frac{\gamma}{1-\overline{q_v}} \; \right] \:
   \overline{\theta} \: \;  \frac{\partial \overline{q_v}}{\partial z}
   \nonumber \;  ,
\end{align}

\vspace{-3mm}
\noindent where the moist-entropy Lewis turbulent number is $\mbox{Le}_{ts} = K_s/K_w$.

If $\mbox{Le}_{ts} = 1$, the second line of (\ref{equ_flux_ths_Lets}) cancels out and $K_h = K_s = K_w$ allows to write the flux of $\theta$ as $\overline{w'\, \theta'} = - \, K_h \: \partial \overline{\theta}/\partial z$, in terms of the exchange coefficient $K_h$.

Differently, if $\mbox{Le}_{ts} \neq 1$, the second line of (\ref{equ_flux_ths_Lets}) exists and the flux of $\theta$ is not proportional to the sole vertical gradient of $\theta$: it also depends on the vertical gradient of $q_v$.
This prevents defining properly an ``exchange coefficient $K_h$ for $\theta$'', and the turbulence must clearly be applied to $\theta_s$ and $q_v$, and not to $\theta_l = \theta$ and $q_v$.

It is thus important to try to determine, from observations and/or from numerical results, whether $\mbox{Le}_{ts} = 1$ or if $\mbox{Le}_{ts}$ is significantly different from unity?

\vspace{-0.5mm}
 \section{\underline{\Large Results}} 
\vspace{-4mm}

\begin{figure}[hbt]
\centering
\includegraphics[width=0.85\linewidth]{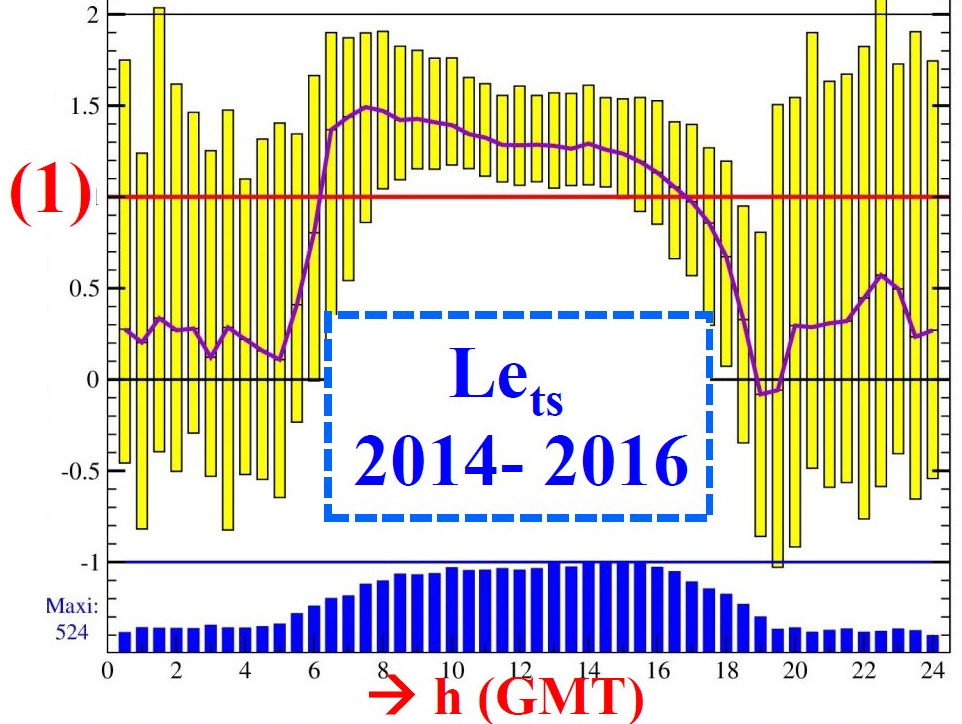}
\includegraphics[width=0.85\linewidth]{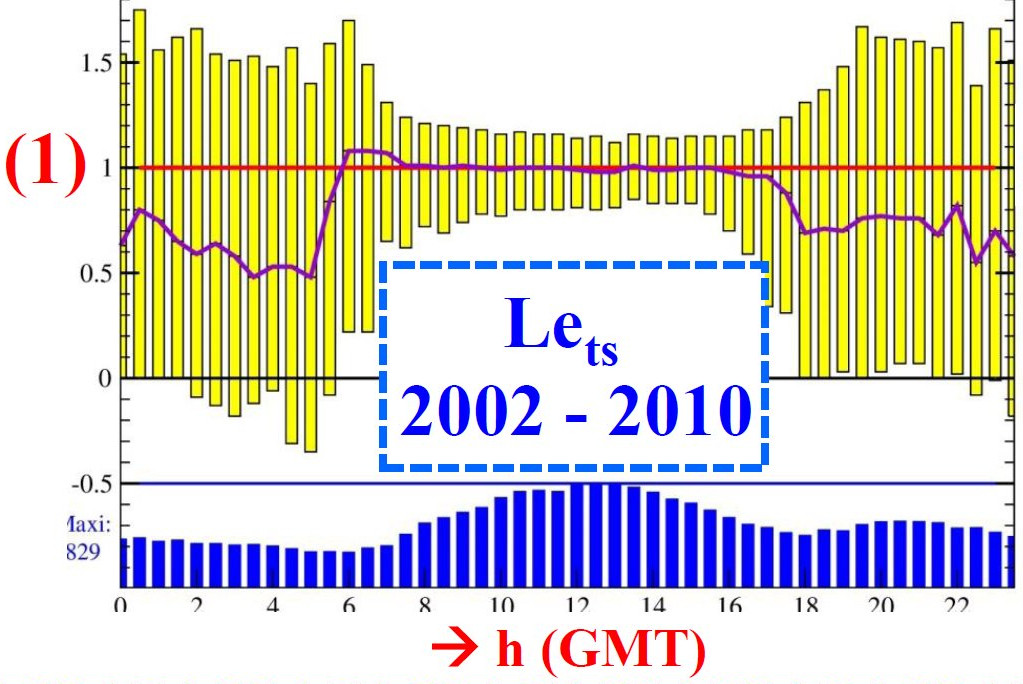}
\vspace{-3mm}
\caption{\small \it 
The boxplots for the Lewis number $\mbox{Le}_{ts} = K_s/K_w$ in terms of the GMT hours.
\underline{Top}: the M\'et\'eopole-Flux mast for $2$ years (CNRM at Toulouse, France).
\underline{Bottom}: thanks to F. Bosvelt, the Cabauw mast for $9$~years (KNMI at Utrecht, Holland).
The number of observations for each class of GMT hours are in blue vertical bars.
The level $\mbox{Le}_{ts} = 1$ is in red and median values are in purple piecewise curve.
\label{fig_1}
}
\vspace{-8mm}
\end{figure}

Figures~\ref{fig_1} show that average yearly Lewis turbulent numbers computed with the the eddy-correlation method are significantly larger than unity in daytime, and are lower than $1.0$ at night.
The significant level is more often reached for monthly averages Figures (not shown) and this diurnal cycle is also observed almost each days, with a maximum present just after sunrise.
This maximum of $\mbox{Le}_{ts}$ is often larger than $2$ in June-August and if often smaller than $0.8$ in winter.

The observed diurnal cycle for $\mbox{Le}_{ts}$ may explain the previous disagreements in the articles of the 1970s: 
values close to $1.0$  may be observed in the late afternoon and values larger than unity in the early morning?

\begin{figure}[hbt]
\centering
\includegraphics[width=0.48\linewidth]{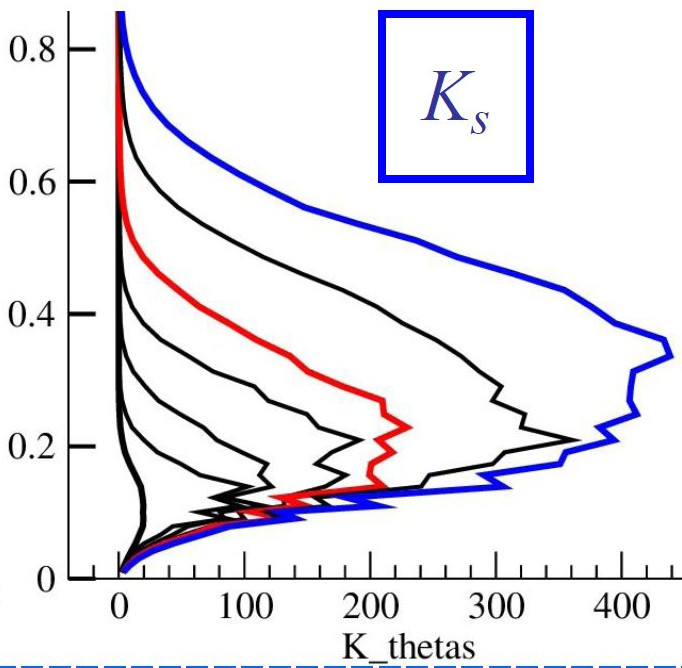}
\includegraphics[width=0.37\linewidth]{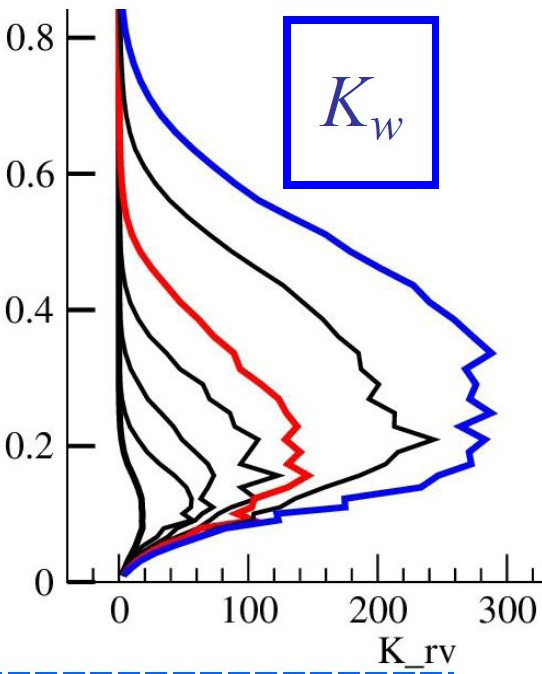}
\includegraphics[width=0.53\linewidth]{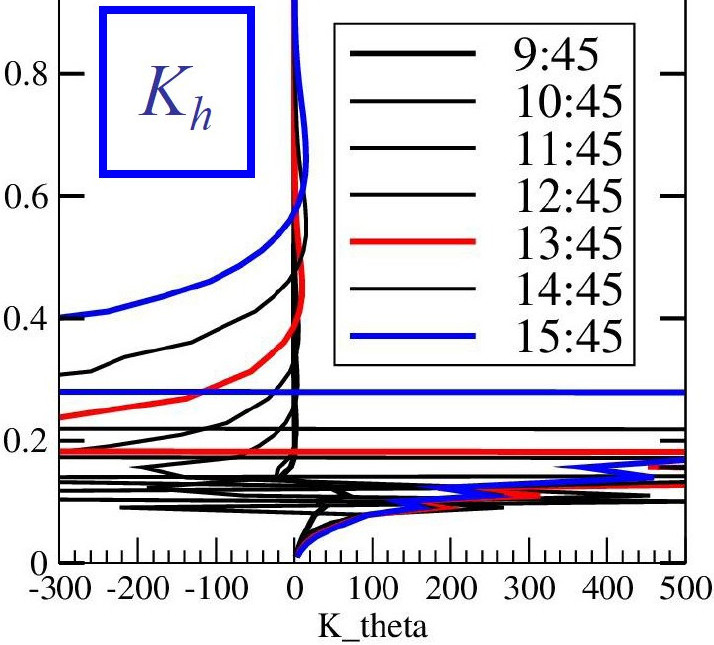}
\includegraphics[width=0.44\linewidth]{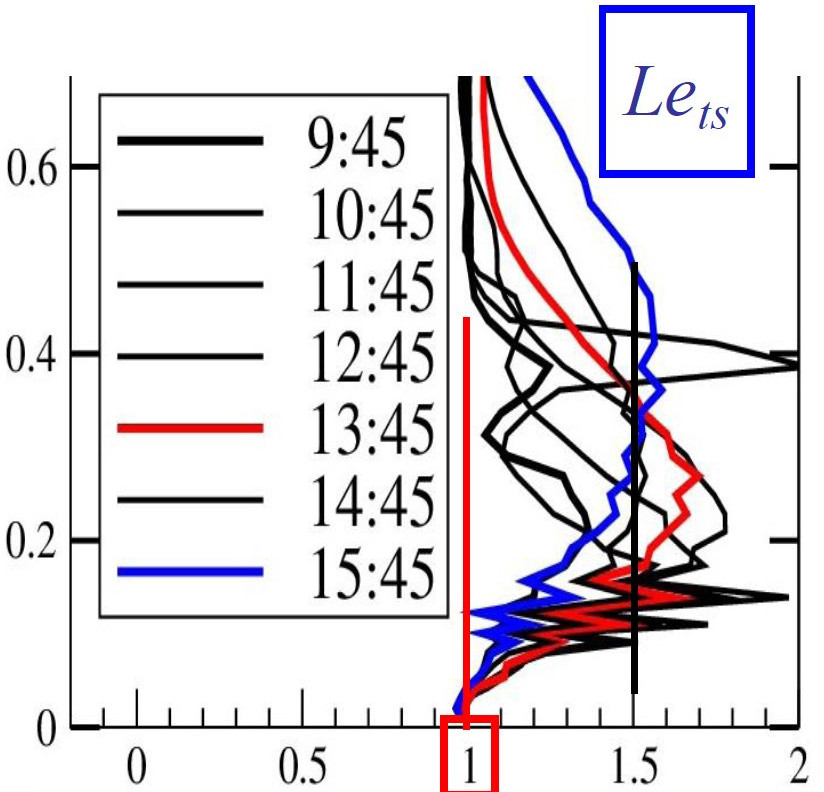}
\vspace{-3mm}
\caption{\small \it 
The vertical profiles of $K_s$, $K_w$, $K_h$ and $\mbox{Le}_{ts} = K_s/K_w$ for hourly averages of a LES run for the moist (but clear-air) case IHOP.
Heights are in ordinates from $0$ to $6$~km or $8$~km.
\label{fig_2}}
\vspace{-6mm}
\end{figure}

Figure~\ref{fig_2} shows that the LES outputs for the IHOP case lead to robust computations of the new moist-air entropy exchange coefficient $K_s = - \, (\overline{w'\, \theta'_s}) / (\partial \overline{\theta_s}/\partial z)$ and of $K_w$ by means of the eddy-correlation method, whereas values of $K_h$ determined from the flux and the vertical gradient of $\theta$ is subject to infinite values (due to zero vertical gradient from about $150$ to $300$~m, depending on the hour) and to a counter-gradient region (due to the same signs of flux and vertical gradient above the level of infinite value).

The turbulent Lewis number is close to $1.0$ close to the surface on Figure~\ref{fig_2}, and then increases with altitude, reaching values above $1.5$ above the $150$~m height where the mass-flux starts to be active.

\vspace{-0.5mm}
 \section{\underline{\Large Conclusion}} 
\vspace{-4mm}

Values of the turbulent Lewis number significantly smaller or larger than $1.0$ are observed for the M\'et\'eopole-Flux and the Cabauw masts, and simulated by a LES of the IHOP case.
Consequently, it is necessary to revisit the equations of turbulence and to determined the values of several new constant coefficients (for the pressure and the second-order moment terms).
These observations and simulations outputs will be used to compute these coefficients.

\vspace{0.5mm}
\noindent{\large\bf \underline {References}}
\vspace{0mm}

{\small \noindent{$\bullet$ Betts AK.} (1973).
Non-precipitating cumulus convection and its parameterization.
{\it Q. J. R. Meteorol. Soc.\/}
{\bf 99} (419):
178--196.
}

\vspace{-0.87mm}{\small \noindent{$\bullet$ Couvreux F. {\it et al.\/}} (2005). 
Water-vapour variability within a convective boundary-layer assessed by large-eddy simulations and IHOP 2002 observations.
{\it Q. J. R. Meteorol. Soc.} 
{\bf 131} (611):
p.2665-2693. 
\url{http://dx.doi.org/10.1256/qj.04.167}
}

\vspace{-0.8mm}{\small \noindent{$\bullet$ Hauf T. and H\"oller H.} (1987).
Entropy and potential temperature. 
{\it J. Atmos. Sci.}, {\bf 44} (20): p.2887-2901.
}

\vspace{-0.8mm}{\small \noindent{$\bullet$ Marquet P.} (2011).
Definition of a moist entropic potential temperature. 
Application to FIRE-I data flights.
{\it Q. J. R. Meteorol. Soc.}
{\bf 137} (656):
p.768--791.
\url{http://arxiv.org/abs/1401.1097}
}

\vspace{-0.8mm}{\small \noindent{$\bullet$ Marquet P.} (2015).
An improved approximation for the moist-air entropy potential temperature $\theta_s$.
{\it WGNE Blue-Book\/}.
\url{http://arxiv.org/abs/1503.02287}
}

\vspace{-0.8mm}{\small \noindent{$\bullet$ Marquet P.} (2016).
The mixed-phase version of moist-air entropy.
{\it WGNE Blue-Book\/}.
\url{http://arxiv.org/abs/1605.04382}
}

\vspace{-0.8mm}
{\small \noindent{$\bullet$ Richardson L. F.} (1919).
Atmospheric stirring measured by precipitation. 
Proc. Roy. Soc. London (A). 96: p.9-18. \url{https://ia600700.us.archive.org/32/items/philtrans07640837/07640837.pdf}
}

  \end{document}